\documentclass[letterpaper]{article} 
\usepackage{aaai25}  
\usepackage{times}  
\usepackage{helvet}  
\usepackage{courier}  
\usepackage[hyphens]{url}  
\usepackage{graphicx} 
\usepackage{natbib}  
\usepackage{caption} 
\usepackage{algorithm}
\usepackage{algorithmic}

\usepackage{newfloat}
\usepackage{listings}
\DeclareCaptionStyle{ruled}{labelfont=normalfont,labelsep=colon,strut=off} 
\floatstyle{ruled}
\newfloat{listing}{tb}{lst}{}
\floatname{listing}{Listing}
\usepackage{xcolor}

\usepackage{multirow}
\usepackage{booktabs}
\usepackage{tabularx}
\usepackage{algorithm}
\usepackage{worldflags}
\flagsdefault[width=2mm]

\usepackage{amsmath,amsfonts}
\usepackage{algorithmic}
\usepackage{graphicx}
\usepackage{textcomp}
\usepackage{bmpsize}
\usepackage{xcolor}
\usepackage{lipsum}
\usepackage{enumitem}
\usepackage[belowskip=2pt]{caption}
\usepackage{newfloat}
\usepackage{longtable}
\usepackage{ltablex}
\usepackage{xltabular}
\newcommand{\paragraphb}[1]{\vspace{0.03in}\noindent{\bf #1.} }

\newenvironment{myquote}%
  {\list{}{\leftmargin=0.0in\rightmargin=0.0in}
  \vspace{-1pt} \item[] \vspace{-2pt}}%
  {\endlist \vspace{-2pt}}
\AtBeginEnvironment{myquote}{\itshape}

\usepackage[most]{tcolorbox}
\DeclareCaptionType{PromptBox}

\usepackage{xcolor}
\newcommand{\answerYes}[1]{\textcolor{blue}{#1}} 
\newcommand{\answerNo}[1]{\textcolor{teal}{#1}} 
\newcommand{\answerNA}[1]{\textcolor{gray}{#1}} 

\definecolor{block-gray}{gray}{0.9}
\newtcolorbox{blockquote}{colback=block-gray,grow to right by=-0mm,grow to left by=-0mm,boxrule=0pt,boxsep=0pt,breakable}

\title{LLM-Powered Analysis of IoT User Reviews: \\ Tracking and Ranking Security and Privacy Concerns}
\author{
    Taufiq Islam Protick\textsuperscript{\rm 1}, Sai Teja Peddinti\textsuperscript{\rm 2}, Nina Taft\textsuperscript{\rm 2}, Anupam Das\textsuperscript{\rm 1}
}
\affiliations{
    \textsuperscript{\rm 1}North Carolina State University, Raleigh, NC, USA\\
    \textsuperscript{\rm 2}Google Inc., Mountain View, CA, USA\\
   tprotic@ncsu.edu, psaiteja@google.com, ninataft@google.com, anupam.das@ncsu.edu
}

\usepackage{bibentry}

\usepackage{fancyhdr}

\begin{document}

\pagestyle{fancy}
\renewcommand{\headrulewidth}{0pt}
\fancyhf{}
\cfoot{\Large{\color{blue}Accepted at AAAI ICWSM 2026}}

\maketitle

\begin{abstract}

Being able to understand the security and privacy (S\&P) concerns of IoT users brings benefits to both developers and users. To learn about users' views, we examine Amazon IoT reviews - one of the biggest IoT markets. This work presents a state-of-the-art methodology to identify and categorize reviews in which users express S\&P concerns. We developed an automated pipeline by fine-tuning GPT-3.5-Turbo to build two models: the Classifier-Rationalizer-Categorizer and the Thematic Mapper. By leveraging dynamic few-shot prompting and the model's large context size, our pipeline achieved over 97\% precision and recall, significantly outperforming keyword-based and classical ML methods. We applied our pipeline to 91K Amazon reviews about fitness trackers, smart speakers and cameras, over multiple years. We found that on average 5\% contained S\&P concerns, while security camera exhibited the highest prevalence at 10\%.  Our method detected significantly more S\&P-relevant reviews than prior works: 15x more for fitness trackers, 29\% more for smart speakers, and 70\% more for cameras. Our longitudinal analysis reveals that concerns like surveillance and data control have persisted for years, suggesting limited industry progress.  We demonstrate that across all device types, users consistently demand more precise control over what data is collected and shared. We uncover challenges in multi-user and multi-device interactions, identifying two previously unreported themes concerning inadequate controls for account separation and data access. These findings, ranging from broad persistent trends to specific instances of customer loss, offer actionable insights for developers to improve user satisfaction and trust.

\end{abstract}

%

\section{Introduction}

The Internet of Things (IoT) market is 
estimated to have 29 billion connected devices by 2027. Popular IoT devices, such as fitness trackers, monitor health metrics; smart speakers manage media, calendars, messaging, and web queries; and security cameras enhance home safety. Despite their utility, these devices pose serious security and privacy (S\&P) risks. Fitness apps may expose location data,
always-listening microphones of smart speakers have led to unauthorized recordings,
smart cameras often lack robust protection against unwanted data access.
Notable breaches like hacks of Foscam and VTech baby monitors underscore these vulnerabilities~\cite{shekyan2013watching}.

S\&P issues raise concerns over data misuse, online fraud, and loss of control which can erode user trust. These concerns are often amplified by media, peer discussions, or unexpected device behaviors~\cite{emami2019exploring}. A UK survey found that security is the third most important factor in IoT purchase decisions, after cost and functionality~\cite{harrisInteractive}. However, S\&P details are often hard to understand before purchase~\cite{emami2019exploring}. To address this, researchers and policymakers have proposed comprehensive S\&P labels~\cite{emami2020ask}. During 2021-2023, the U.S. government advanced this effort through NIST, proposing a Cyber Trust mark with a QR code linking to detailed information~\cite{chen2024trustmark}. Such labels are not yet widely adopted, and similar privacy labels in mobile apps remain insufficient~\cite{zhang2023privacy}.

Advances in natural language processing (NLP) now allow automated analysis of product reviews at scale~\cite{harkous2022hark},  surpassing traditional machine learning in quality. These techniques can extract S\&P concerns across devices, offering value to consumers, developers, and researchers. For consumers, this information supports informed purchase decisions. Developers can use feedback to improve product security, prioritize issues, and address root causes of negative reviews. Notably, 60\% of S\&P-related reviews in one study led to relevant app updates~\cite{nguyen2019short}.
Retailers have started integrating AI-powered shopping assistants. For instance, Amazon’s “Rufus,” launched in September 2024, allows users to query product details and review summaries. However, it currently overgeneralizes feedback and lacks focus on S\&P concerns. Our work aims to enhance such assistants by improving their ability to identify and summarize S\&P issues effectively.

Prior research on users' S\&P perceptions has relied on small-scale surveys~\cite{emami2021privacy} or keyword-based analyses~\cite{protick2024unveiling}. The most advanced prior work~\cite{harkous2022hark} used a T5 model and showed improved accuracy over manual and classical methods like LDA or SVMs (for apps, not IoT products). To our knowledge, we are the first to leverage LLMs for this task.

Our first research question asks \textbf{RQ1: \textit{What level of quality can be achieved for LLM-based detectors aimed at extracting S\&P reviews from a large corpus of IoT reviews?}} We aim to understand if one detector can uncover a wide range of S\&P issues, while simultaneously reaching high accuracy. With the emergence of shopping assistants, it is important to understand the quality of such a detector and its ability to uncover a wide range of issues. While we expect LLMs to do well, our work offers a benchmark for future efforts since the labeled data we use is public. We fine-tune GPT-3.5-Turbo to build two models: the \emph{Classifier-Rationalizer-Categorizer} (CRC), which detects and labels S\&P issues in reviews, and the \emph{Thematic Mapper} (TM), which groups these issues into higher-level themes based on established taxonomies. We see that our models identify substantially more S\&P-relevant reviews than prior approaches.

With our models, we explore two more research questions.
\textbf{RQ2: \textit{What is the prevalence of S\&P concerns in IoT product reviews, and how does prevalence vary across device categories?}} 
We analyze Amazon reviews of fitness trackers, smart speakers, and security cameras to measure how frequently users raise S\&P concerns and how this varies by device type.
\textbf{RQ3: \textit{What are users' S\&P concerns and how do they rank relative to one another?}}
Our LLM-based pipeline supports issue extraction and ranking, allowing for trend analysis of S\&P issues across device types over time.

In summary, we make the following contributions: 

\begin{itemize} [noitemsep,nolistsep,topsep=2pt,leftmargin=1.em]
    \item We present a state-of-the-art approach by fine-tuning GPT-3.5-Turbo to develop two models. The first model, \textit{CRC}, achieves over 97\% precision and recall on a human-annotated validation set, and macro-averaged precision and recall of 88.7\% and 93.4\%, respectively, on real-world data. The second model, \textit{TM}, achieves a micro-averaged precision of 96.8\%, recall of 95.7\%, and an F1-score of 96.3\%. Comparative analysis with existing approaches demonstrates that our method outperforms traditional keyword-based techniques, classical ML/NLP models, and baseline LLMs.
    
    \item We applied our pipeline to 91K IoT product reviews and identified 4,896 reviews ($\approx5\%$) with S\&P concerns. Security cameras had the highest rate of such concerns (10.07\%), significantly more than fitness trackers and smart speakers. We also observed a slight negative correlation between S\&P mentions and star ratings, suggesting these concerns may contribute to lower satisfaction.

    \item The combination of more accurate methods (that extract more content than prior efforts), coupled with our categorizations of S\&P concerns into 28 categories (permitting more distinction than prior work \cite{vetrivelexamining} with 7 categories) enables simultaneously a broad view across issues (top issues, issue ranking, temporal trends), and also specific insights (e.g., trust, ethics, customer loss, and the variety of specific controls users seek). Surveillance dominates concerns for smart speakers and cameras. For both fitness trackers and cameras, the top three issues persist over time, indicating limited progress in addressing user concerns. A common, cross-device theme is users' desire for finer control over both data types and recipients, including household members, visitors, landlords, corporations, and government agencies.
    
    \item Our analysis further uncovers significant challenges in multi-user interactions with IoT devices, identifying two previously unreported themes that reveal distinctive privacy risks associated with cross-device interactions. Users frequently report inadequate and poorly designed controls for account separation and data access. Examining 52 speaker and camera device types over multiple years, our study highlights the widespread nature of these issues. We conclude by summarizing these challenges as actionable feedback for developers.

\end{itemize}

\begin{table*}[!t]
\resizebox{1.0\textwidth}{!}{
\begin{tabular}{c c c c c c c} 
\toprule
\textbf{Context} & \textbf{Reference} & \textbf{App/Device-Types (\#)} & \textbf{Training Data} & \textbf{Model} & \textbf{\% of S\&P reviews (\#)} & \textbf{S\&P Themes}\\

\midrule

\multirow{4}{*}{Mobile} & \cite{cen2014user} & 6,938 apps & keywords & Logistic Regression & 0.05\% (5M) & 5 \\
\cline{2-7}
\multirow{4}{*}{Apps} & \cite{nguyen2019short} & 2,583 apps & keywords & SVM & 0.12\% (4.5M) & N/A\\
\cline{2-7}
\multirow{4}{*}{Reviews} & \cite{mukherjee2020empirical} & 539 apps & keyword & SVM  & 0.5\% (2.2M) & N/A\\
\cline{2-7}
 & \cite{besmer2020investigating} & 5M apps & keyword & Logistic Regression & 0.07\% (4.9M) & 9\\
\cline{2-7}
 & \cite{harkous2022hark} & 1.3M apps & taxonomy+NLI & T5 & 0.9\% (626M) & 25 (P only) \\
\cline{2-7}
 & \cite{akgul2024decade} & 160K apps (over 10 yrs) & taxonomy+NLI & T5  &  0.65\% (1.9B) & 89 (P only)\\

\midrule  

\multirow{3}{*}{IoT} & \cite{fruchter2018consumer} & speakers (5) & keywords & Manual & 2.0\% (109K) & 3\\
\cline{2-7}
\multirow{3}{*}{Product} & \cite{van2020pets} & activity \& location trackers (14) & keywords & Manual & 0.14\% (20K) & 8\\
\cline{2-7}
\multirow{3}{*}{Reviews}& \cite{vetrivelexamining} & cameras (881), routers (440), hubs (54), set-top boxes (41)  & keywords & LDA  & 9.8\% (68K) & 22\\
\cline{2-7}
& \cite{protick2024unveiling} & trackers (3), speakers (3), cameras (7) & keywords & Manual & N/A & 14\\
\cline{2-7}
 & \textbf{Our Work} & trackers (23), speakers (24), cameras (28) & qualitative & GPT-3.5-Turbo & 5\% (91K) & 28\\

\bottomrule

\end{tabular}}
\caption{Comparison with existing works that elicit security and/or privacy issues from user generated reviews.}
\label{tab:comparison-prior-work}
\vspace{-16pt}
\end{table*}

\section{Related Work}
\paragraphb{Analysis of Mobile 
App Reviews}
Multiple studies have analyzed Google Play reviews to identify S\&P concerns in mobile apps. Nguyen et al. use an SVM classifier to detect privacy-relevant reviews and correlate them with app updates~\cite{nguyen2019short}, while Mukherjee et al. apply a similar approach to study S\&P concerns~\cite{mukherjee2020empirical}. Cen et al. perform two-dimensional labeling of S\&P issues using supervised ML~\cite{cen2014user}. Besmer et al. combine ML and NLP to explore privacy perceptions based on reviews and ratings~\cite{besmer2020investigating}. Harkous et al. use deep learning to summarize privacy feedback~\cite{harkous2022hark}, and Akgul et al. analyze 12 million privacy-relevant reviews using a large language model~\cite{akgul2024decade}. Our work builds on these efforts by leveraging a more advanced LLM. These latter 2 works focused on app reviews, while we study IoT reviews.  

\paragraphb{Analysis of IoT Product Reviews}

Linden et al. analyzed 20K reviews of human wearables (e.g., fitness trackers) and found privacy concerns were rarely expressed (0.14\%)~\cite{van2020pets}. Fruchter et al. used a 53-keyword lexicon to identify S\&P concerns in 109K smart speaker reviews, finding 2.04\% raised issues related to data collection, creepy behavior, and privacy violations~\cite{fruchter2018consumer}. Vetrivel et al. examined 68K Amazon reviews of surveillance systems, hubs, set-top boxes, and routers, identifying S\&P concerns in 9.8\% of them~\cite{vetrivelexamining}; for cameras specifically, they report only 5.9\% compared to our 10.07\%. Most recently, Protick et al. used a keyword-based approach to qualitatively analyze smart home IoT reviews, surfacing common and unique S\&P issues, such as invasive customer support and user skepticism~\cite{protick2024unveiling}.

\paragraphb{Comparison with Prior Works} 
To our knowledge, we are the first to fine-tune a large language model on unstructured IoT product reviews. Prior studies~\cite{van2020pets,fruchter2018consumer,protick2024unveiling} typically rely on keyword-based filtering, which struggles to capture semantically similar content and is prone to high false positive and false negative rates~\cite{harkous2022hark}.
While large open-source models like T5 have been used to analyze privacy-related mobile app reviews~\cite{harkous2022hark,akgul2024decade}, our method leverages a (larger size~\footnote{T5-11B has 11 billion parameters. The exact size of GPT-3.5-Turbo is unknown, but is estimated to be larger than T5-11B.}) closed-source API-based model that does not need self-hosting for fine-tuning and is continuously updated. Also, we focus on the IoT ecosystem, such as fitness trackers, smart speakers, and cameras—that involve distinct data modalities and S\&P risks.
Moreover, previous methods often rely on hypothesis-based Natural Language Inference (NLI) to bootstrap data for manual labeling and later use it for fine-tuning large models. In contrast, we transform a pre-existing IoT labeled dataset~\cite{protick2024unveiling}, and fine-tune GPT-3.5-Turbo using dynamic few-shot prompting that effectively utilizes the larger context size of 16K tokens by including manually labeled examples.
And lastly, while prior work has utilized four models/steps to organize reviews into thematic topics (one fine-tuned model for privacy-relevant review detection, another fine-tuned model for low-level issue tag generation, leader clustering technique to merge these low-level issues into higher level themes, and one fine-tuned generative model to generate theme titles), we compressed these into two fine-tuned models (CRC and TM) and focused the thematic analysis on 28 high-level themes derived from a unified S\&P taxonomy merged from eight vetted IoT taxonomies. Our methodology requires fewer models and is anchored on vetted taxonomies.
In summary, our study is not keyword-dependent, covers 28 nuanced S\&P themes (e.g., privacy ethics, data accuracy), and enables automated analysis of IoT product reviews over time. This approach captures both the evolution and ranking of issues, offering actionable insights to help developers prioritize mitigations. Table~\ref{tab:comparison-prior-work} summarizes these key differences with existing related works.

\section{Methodology}

\subsection{Data Description}

\paragraphb{Data Collection}
We adopted the following steps for data collection.
First, we go to \texttt{amazon.com} and search for device categories like ``fitness trackers", ``smart home speakers", and ``security cameras". 
We focus on these three device categories for two key reasons. First, they encompass diverse data modalities --- such as motion sensors, heart rate, GPS, audio, and video --- each presenting distinct security and privacy risks. Second, this selection allowed us to leverage thousands of existing human-annotated data points from existing work~\cite{protick2024unveiling} for model training and evaluation.
We next list products and their ASINs from the first search page. Then, with a crawler created with Selenium and BeautifulSoup, we collect review texts and metadata (including rating, user's country information, date, and ASIN product IDs). The data was gathered in May 2023. Finally, we preprocess the crawled data into the wild review dataset ($\mathcal{W}$), described in more detail later in the section. 

\paragraphb{Ethical Considerations}
\label{sec:ethics}
We comply with Amazon's terms and conditions of use during our data collection. Our data crawling process strictly follows the guidelines outlined in Amazon's \texttt{robots.txt} file, particularly when accessing product review pages. Furthermore, we implement rate limiting for our crawlers to minimize server load and avoid any adverse impact on their infrastructure. To safeguard consumers' privacy, we exclude any personally identifiable information, even if publicly accessible, such as usernames and posted images, as part of our data collection process. 
Also, fake reviews are a prominent problem on online marketplaces (including on Amazon)~\cite{he2022market}, and we rely on Amazon's mechanisms for removing fake reviews and banning abusive accounts.
Since we are collecting publicly available data, our IRB office, upon review, classified the study as an exempt.

\paragraphb{Wild Review Dataset ($\mathcal{W}$)} \label{thewilddataset}
The crawled Amazon reviews are preprocessed to obtain a good quality wild review dataset $\mathcal{W}$. We collected a total of 127,821 reviews, which, after removing the empty entries, came down to 127,270 reviews. The number came down to 114,383 reviews after filtering out non-English content. We then eliminated duplicate reviews, resulting in 91,749 reviews. 
The distribution of reviews across the three device categories is as follows: trackers (24,046), speakers (32,179), and cameras (35,524). 
These reviews are for products from 75 vendors (trackers: 23; speakers: 24; cameras: 28), identified by ASINs. We will be making this wild review dataset public. With Amazon changing its policy towards the end of our data collection period and drastically reducing the number of reviews that can be collected (from 5000 to just 100), we hope other researchers will find our dataset useful. The anonymous repository for product details can be found here \footnote{https://anonymous.4open.science/r/IoT-Products-SecurityAndPrivacy-5F5D/README.md}.

\paragraphb{Qualitative Labeled Data for Training}
\label{sec:qualitative-data}

We fine-tuned our models using pre-labeled qualitative data from previous research eliciting S\&P concerns from IoT product reviews \cite{protick2024unveiling}. The dataset follows a three-step analysis of IoT products: keyword selection, review labeling, and thematic analysis, all conducted by multiple coders with high inter-rater reliability. 

\subsection{Model Choice \& Rationale for Fine-tuning}
To address \textbf{RQ1} in evaluating whether LLM-based detectors can effectively extract S\&P reviews, we opted to use the GPT-3.5-Turbo model.
We choose GPT-3.5-Turbo due to its high throughput limits (up to 1M tokens/min, 100M tokens/day for Tier-4 users), cost-efficiency, and strong performance across tasks. The computation was done on a machine that uses Linux with 16 GB of RAM. It uses an NVIDIA RTX-3070-M GPU for faster computation of BERT embedding. For deterministic outputs, we set the \texttt{temperature} parameter to 0, and to overcome hallucination-induced deviations in the LLM response, we set both \texttt{frequency\_penalty} and \texttt{presence\_penalty} values to 2. 

We utilize the Supervised fine-tuning (SFT) method to adapt the GPT-3.5-Turbo model for our case. Rather than simply give input-output pairs during fine-tuning, we leveraged the larger context size of the model (16K tokens) to include additional few-shot examples that are dynamically chosen for each input text in the training data. 
We fine-tune the base model for multiple reasons, all of which are related to the downstream task of our interest. First, accurate identification of S\&P concerns requires domain knowledge, which the base model may lack without fine-tuning (we provide a comparative analysis of vanilla vs. fine-tuned models in the next section). Second, the base model exhibits inconsistent responses for the same input, even with the \texttt{temperature} set to 0. Finally, it often fails to follow the structured response format required for our task.

Although GPT-3.5-Turbo is sometimes described as a “base model,” it is already instruction-tuned and not a raw pre-trained foundation model. Our work adds an additional round of task-specific supervised fine-tuning to adapt it to IoT security and privacy concerns. For fine-tuning, we use OpenAI’s API with 3 epochs and evaluate temperatures between 0.0 and 0.8 in increments of 0.2. The model's learning rate is internal. However, the API gives a learning rate multiplier value to be used, which in our case was set to 2. The 16K-token context window allows dynamic few-shot examples. The best performance is achieved at temperature = 0.

\subsection{Classifier-Rationalizer-Categorizer Model}
\label{sec:crc-section}
Given a dataset of reviews, our first goal is to detect S\&P concerns mentioned in these reviews. For this, we focus on three key aspects: first, determining whether a review text contains any concerns (\emph{classify)}; second, obtaining a justification/explanation for the decision (\emph{rationalize}); and third, determining what specific concerns are present in the review text (\emph{categorize} into low-level issues). We have modeled these three aspects as three tasks to be performed on a given review text `$t$', as elaborated in this section.
We have included the second step based on prior work showing that `chain-of-thought' prompting improves performance on a range of reasoning tasks~\cite{wei2022chain}. 

\paragraphb{Dataset for Training and Validation}
\label{sec:trainingdata}
We leverage an existing dataset from \cite{protick2024unveiling} containing IoT product reviews labeled for S\&P concerns. However, this dataset was created for a different purpose and has different semantics. So we first manually transform this dataset according to our task-set requirements, as shown in Table~\ref{need-crc} in Appendix \ref{tmappdx}. Since this pre-existing dataset was already labeled by two annotators and went through rigorous disagreement resolution to generate high-quality labels, one of the authors focused on manually transforming this data into our desired format. The labels for task-1 (\textit{classify}) and task-3 (\textit{categorize}) were directly derived from prior labels, where reviews containing S\&P concerns were considered as positive samples and the rest as negative. The same author manually augmented these labels by generating explanation strings (i.e., task-2 \textit{rationalize} output) to capture the reasons behind these labels based on prior annotations. As we focused on data transformation rather than new-label generation, multiple annotators were not involved and inter-annotator agreement metrics were not computed.
Since there were not enough negative samples in this pre-existing dataset, we augmented it with additional negative samples from our crawled dataset, $\mathcal{W}$, through manual selection until approximately 600 were identified (this reduced our $\mathcal{W}$ dataset from 91,749 to 91,149 reviews). Our final labeled dataset contains 2,454 reviews, \emph{equally divided} between positive and negative samples.

\paragraphb{Prompt Engineering} \label{subsec:prompt}
Given the complexity of the CRC task (containing 3 sub-tasks), we curate a good prompt for conducting Supervised fine-tuning (SFT) of the GPT-3.5-Turbo model that takes advantage of its longer context window (16K tokens). We utilize the knowledge gained in the manual data transformation process to create this prompt. For a given review text 
and its converted labels (right column of Table \ref{need-crc}), we finalized the prompt shown in Figure~\ref{fg:A}, where underlined texts vary for each input review text.

\begin{figure}[!t]
\fbox{
\footnotesize
\begin{minipage}{.97\columnwidth}
------ System Prompt ------\\
You are a Large Language Model for classifying, rationalizing, and categorizing security and privacy concerns in text.\\
------ User Prompt ------\\
Given the text t: \underline{\texttt{$t$, a review to label}}\\
I want you to perform three tasks on t:\\
Task 1: Determine if text t explicitly mentions any security or privacy concerns. Respond only with a `Yes' or a `No'.\\
Task 2: Provide a brief explanation as to why your response is as such for Task 1.\\
Task 3: If the response for Task 1 was a `Yes', categorize the explicit concerns mentioned in text T with themes related to security and privacy.\\
    \textit{\textbf{[}At this point, a random number $x$ is generated between 0 and 1. If $x < 0.5$, we assign Example A = $t_{1}$, the text closest to $t$. Else, we assign Example A = $t_{2}$, the text second closest to $t$ with the opposing label as that of $t_{1}$ from the training dataset\textbf{]}}\\
    Here are two examples that might be helpful:\\
    Example text A: \underline{\texttt{$x$ < 0.5 ? $t_{1}$: $t_{2}$}}\\
    Ideal response for A: \underline{\texttt{$x$ < 0.5 ? $Label(t_{1})$: $Label(t_{2})$}}\\
    Example text B: \underline{\texttt{Show the other example}}\\
    Ideal response for B: \underline{\texttt{Show the other example's label}}
  \end{minipage}}
  \caption{The prompt for the CRC model}
\label{fg:A} 
\vspace{-4pt}
\end{figure}

The first statement is the system prompt, which guides the model's behavior. What follows is the user prompt, where we first give a review text $t$ and specify the task set that allows the model to know what to do with $t$. We show two helpful examples $t_{1}$ and $t_{2}$ with their ideal responses from the training dataset so that the generative model can understand the correctness and the desired response format for $t$. Instead of fixing these few-shot examples in the prompt, we dynamically choose these examples (dynamic few-shot prompting) based on $t$. We select review text $t_{1}$ that is closest to $t$ in the embedding space, and $t_{2}$ that is second closest to $t$ in the embedding space but has an opposite label for task-1. Here, the texts are represented with their BERT embeddings~\cite{kenton2019bert} (which help represent text in a multi-dimensional space, where semantically related texts are closer together) and closeness is calculated with cosine similarity. 
To make sure we do not have label-specific bias, 
$t_{2}$ is chosen to have the opposite task-1 label from $t_{1}$ (Line 18 of Algorithm \ref{euclid} in Appendix \ref{sec:algos}), i.e., $t_{2}$ is a helpful review with no S\&P concerns if $t_{1}$ is a review with S\&P concerns, and vice versa. 
To further reduce positional bias between the examples, 
we randomly show the closest or the second closest example first based on a coin toss (text in square brackets in Figure~\ref{fg:A}). Note that $Label(t_i)$ includes the outputs for all three tasks for $t_i$ (classification, rationale, and categorization outputs). An example of a CRC prompt with all values filled in is shown in Appendix \ref{crctemplateexample}.

\begin{figure*}[!t]
\centering
\includegraphics[width=.9\textwidth]{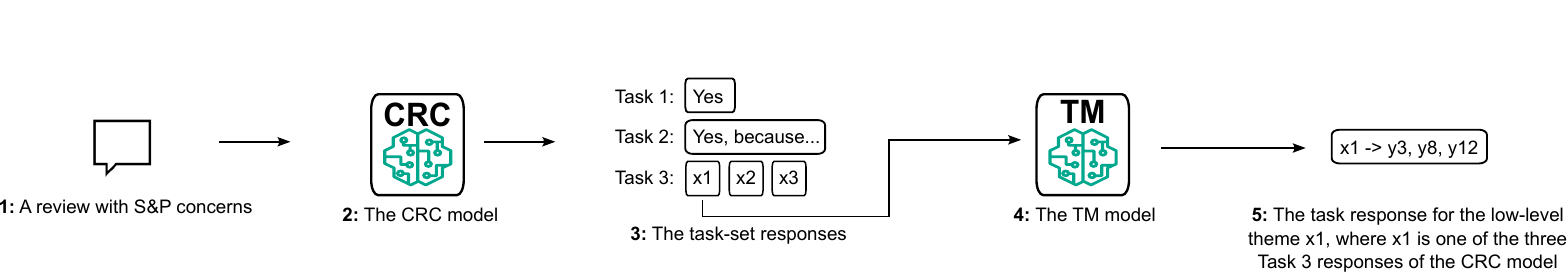}
\caption{The CRC and TM models in tandem. The Task 3 responses of the CRC model are passed to the TM model as input.}
\label{TM_CRC}
\vspace{-12pt}
\end{figure*}

\paragraphb{Model Training} 
During the training phase, we utilize the manually annotated task set responses for each review $t$. 

The model is trained using 80\% of the dataset (1964 reviews), with the remaining 20\% (490 reviews) reserved for validation. The temperature setting ($temp$) is a hyperparameter during model training, and we vary it from $0$ to $0.8$ in increments of $0.2$, resulting in five finetuned models. Each of these models is evaluated on the manually labeled validation set (balanced between positive and negative samples) to identify the best-performing one. We found that the model with $temp=0$ yielded the highest accuracy, precision, recall and F1-score -- all close to 97.8\%. 

\paragraphb{Impact of Prompt Engineering}
To quantitatively assess the utility of the dynamic few-shot prompting, we evaluated the performance of the Vanilla model (i.e., without finetuning) using all three prompting strategies: zero-shot (prompt in Figure~\ref{fg:A} with no examples), fixed-shot (prompt with two fixed examples), and our dynamic few-shot prompting. We found that the dynamic few-shot prompting method achieved the highest performance (89\%) on the validation dataset, outperforming fixed-shot (88\%) and zero-shot (83\%). Our fine-tuned CRC model employing dynamic few-shot prompting further boosted the accuracy to 98\%.

We would like to point out that the closest neighbors identification in the dynamic few-shot prompt is necessary only during training, and we do not actually need to add few-shot examples in the prompt during inference -- the high performance of the model is maintained even if we skip populating the few-shot examples during inference. Hence, the additional cost of computing BERT embeddings and cosine similarities for finding the nearest neighbors is a one-time training cost -- where we compute and store embeddings for all the samples in the training data, and perform quick matrix multiplications to identify nearest neighbors (both positive and negative cases) for each sample. 

\paragraphb{Comparison with Other Models}
\begin{table}[!t]
\scriptsize
\centering
\begin{tabular}{@{}llllll@{}}
\toprule
\textbf{Model}                                                               & \textbf{Detail}          & \textbf{\begin{tabular}[c]{@{}l@{}}Acc. \%\end{tabular}} & \textbf{\begin{tabular}[c]{@{}l@{}}P \%\end{tabular}} & \textbf{\begin{tabular}[c]{@{}l@{}}R \%\end{tabular}} & \textbf{\begin{tabular}[c]{@{}l@{}}F1 \%\end{tabular}} \\ \midrule
\multirow{7}{*}{\begin{tabular}[c]{@{}l@{}}Keyword-\\ based\end{tabular}}    & \cite{cen2014user}              & 71.0                                                       & 92.6                                                    & 45.7                                                    & 61.2                                                     \\
                                                                             & \cite{nguyen2019short}           & 58.0                                                       & 64.4                                                    & 35.5                                                    & 45.8                                                     \\
                                                                             & \cite{mukherjee2020empirical}        & 76.1                                                       & 96.4                                                    & 54.3                                                    & 69.5                                                     \\
                                                                             & \cite{fruchter2018consumer}         & 79.6                                                       & 95.0                                                    & 62.4                                                    & 75.4                                                     \\
                                                                             & \cite{van2020pets}          & 55.5                                                       & 93.5                                                    & 11.8                                                    & 21.0                                                     \\
                                                                             & \cite{vetrivelexamining}         & 78.8                                                       & 86.9                                                    & 67.8                                                    & 76.1                                                     \\
                                                                             & \cite{protick2024unveiling}          & 85.9                                                       & \textbf{98.9}                                           & 72.7                                                    & 83.8                                                     \\ \midrule
\multirow{3}{*}{\begin{tabular}[c]{@{}l@{}}Classical \\ Models\end{tabular}} & Logistic Regression      & 90.4                                                       & 90.2                                                    & 90.6                                                    & 90.4                                                     \\
                                                                             & SVM                      & 84.3                                                       & 80.9                                                    & 89.8                                                    & 85.1                                                     \\
                                                                             & KNN (k = 37)             & 80.2                                                       & 76.1                                                    & 91.0                                                    & 82.9                                                     \\ \midrule
LLM                                                                          & Vanilla GPT-3.5-Turbo          & 88.7                                                       & 83.4                                                    & 96.0                                                    & 89.3                                                     \\
                                                                             & \textbf{CRC (our model)} & \textbf{97.8}                                              & 97.6                                                    & \textbf{98.0}                                           & \textbf{97.8}                                            \\ \bottomrule
\end{tabular}
\caption{Performance comparison between our model and others. Model performance for Class 1 (S\&P concerns), showing precision, recall, and F1-scores.}
\label{tab:performance}
\end{table}
Using the validation dataset of 490 reviews (balanced 245 positive and 245 negative samples), our evaluation (as shown in Table~\ref{tab:performance}) highlights the limitations of keyword-based and classical ML methods as well as the advantages of advanced models like CRC. Keyword-based approaches achieve high precision by relying on fixed terms but suffer from very low recall, missing concerns expressed in varied wording. While this validation dataset is balanced, real data is heavily imbalanced with very few S\&P examples. To evaluate how keyword-based approaches work on real datasets, we curated a second validation dataset containing a random sample of 300 reviews that are manually labeled. This dataset contains eight S\&P and the rest non S\&P reviews. All the keyword-based models have a really high number of false positives, resulting in precisions less than 20\% and recall values less than 50\%.

Classical machine learning models outperform keyword-based approaches by capturing broader patterns and improving recall, but they still face challenges in balancing precision and recall effectively. Prior works using LLMs such as T5~\cite{harkous2022hark,akgul2024decade} have not released their models publicly, preventing direct comparison on the same datasets. In contrast, general-purpose LLMs like Vanilla GPT-3.5 achieve high recall but exhibit lower precision, frequently misclassifying benign reviews as security and privacy concerns. In contrast, our CRC model achieves both precision and recall above 97\%, effectively capturing nearly all true S\&P concerns while minimizing false alarms---a balance critical for trustworthy S\&P analysis. Moreover, vanilla GPT-3.5-Turbo failed to generate outputs for 9\% of validation samples (44 of 490) and achieved only 88.7\% accuracy on the remainder, with 52 misclassifications---five times higher than CRC. By comparison, our fine-tuned CRC model reached 97.8\% accuracy, demonstrating the effectiveness of task-specific fine-tuning in achieving reliable performance.

\paragraphb{Sampled Performance on Wild Dataset} As a sanity check we also randomly sampled 300 reviews from the wild dataset $\mathcal{W}$ (with 100 from each device category), manually labeled them, and used these to verify the model's performance. 

Despite significant class imbalance (only 8 of the 300 samples are positive examples), the model achieves a good performance: 99\% accuracy with macro average precision of 88.7\%, recall of 93.4\%, and F1-score of 90.9\% (weighted average precision,  recall, and F1-score are all 99\%). There were only 2 false positives and 1 false negative, indicating high accuracy and reliability in classifying S\&P concerns. Table \ref{performance} in Appendix~\ref{crctemplateexample} shows the full evaluation of the CRC model across the three different tasks; different metrics are used for evaluating closed and open-ended responses.

\paragraphb{Processing the Full Wild Dataset} Next, we processed the entire wild dataset $\mathcal{W}$ through the model. It took approximately 30 hours to submit all the prompts to OpenAI's API, yielding labels for 91,149 reviews. 

\subsection{Thematic Mapper Model}\label{thetmmodel}

The output of the CRC model helps us quantify the number of S\&P-related reviews and identify the low-level issues raised in these reviews. However, these identified concerns were often too review-specific and contained variability in their expression, yielding too many uniquely worded low-level issues that are similar in nature. This prevents us from succinctly summarizing the key issues users share in their reviews. So we propose to fine-tune a Thematic Mapper (TM) model that will map the low-level issues (i.e., task-3 responses) of the CRC model to one or more high-level themes related to S\&P topics. For instance, both `concerns about eavesdropping' and `concerns about listening' can be mapped to high-level `surveillance' S\&P topic. Such high-level themes are more digestible for stakeholders.

\paragraphb{Collection of High-level Themes}
To ensure broad and comprehensive coverage of S\&P topics, we merged eight vetted IoT taxonomies \cite{williams2019security,rizvi2018securing,alsamani2018taxonomy,das2018taxonomy,gulzar2019internet,akbar2021multicriteria,chen2018internet}. Existing taxonomies for IoT products face two main limitations: (1) some include a limited number of subcategories, omitting important issues like anonymity, privacy ethics, or data accuracy; and (2) some categories are overly broad (e.g., ``user fears,'' could encompass issues like unintended data sharing, account management difficulties, or hacking), making them unhelpful for users to interpret and too ambiguous for developers to know what to fix.
Three S\&P experts from academia and industry collaboratively merged the taxonomies, removing redundancies and disaggregating overly broad topics. All disagreements were discussed and resolved through consensus, consistent with Delphi methodology~\cite{turoff2002delphi}, which emphasizes convergence over inter-rater reliability. The definitions of our 28 themes are available in our anonymous open-source repository ~\cite{openscience}.

\paragraphb{Dataset for Training and Validation}
\label{sec:tmdataset}
To train the TM model, we leverage 1,200 \emph{unique} low-level issues (from 1,989 total) stemming from the positive reviews in the training data for the CRC model, and manually map them to the 28 high-level themes. 
Annotators performed this manual mapping in multiple rounds and resolved any discrepancies through discussion until they arrived at a final agreement. This mapped dataset is used to fine-tune our TM model.


\paragraphb{Prompt engineering}\label{promptenggTM}
Figure \ref{fg:tmpromptengineering} in Appendix \ref{tmappdx} shows the prompt we use to fine-tune the GPT-3.5-Turbo model to be our thematic mapper (the underlined texts vary for each example).
Like in the CRC prompt, the first statement is a system prompt, and this is followed by the user prompt. In the user prompt, we first give the low-level issue $x$ to map. Next, we provide the list of taxonomy terms $Y$ with their definitions, and later follow this with the task that the model should perform on $x$. Finally, we show five helpful examples $t_{1}$ through $t_{5}$ with their correct labels $Label(t_{1})$ through $Label(t_{5})$ from the training dataset, so that the generative model can learn from some ground truth. Similar to the CRC model, we follow a dynamic few-shot prompting technique where the shown examples are chosen based on how close they are to $x$ in the embedding space using cosine similarity. For the TM model, we choose TF-IDF embeddings (Line 3 of Algorithm \ref{euclidtm} in Appendix \ref{sec:algos}) as opposed to BERT embeddings based on its performance on short phrases (like our low-level issues).

For example, when using TF-IDF embedding, the closest helpful example retrieved for `password sharing as a violation of basic IT security principles' is `password security', whereas it is `unauthorized listening or recording in a joking tone' when using BERT embedding.

\paragraphb{Model Training}
Each low-level issue from Task-3 responses from the CRC model is passed to the TM model (see Figure \ref{TM_CRC}). We utilize 70\% of the 1,200 curated low-level to high-level theme mappings 
as our training dataset, and the rest (i.e., 360 or 30\%) as a validation dataset to train our TM model. 
Based on our experience with the CRC model, we set the \texttt{temperature} parameter to 0. To assess the performance of the TM model, we verified each high-level theme prediction made for the 360 low-level issues in our validation dataset. Note that a single low-level theme can be mapped to multiple high-level themes. For instance, `concerns about sharing login details' was mapped to `authentication' and `data sharing'. Two contingency tables of dimension $rows \times terms = 360 \times 28$, one for TM predictions and the other for ground truths, were created for comparison. The model achieves a macro-averaged precision of 90.99\%, recall of 88.14\%, and F1-score of 89.33 (micro-averaged precision is 96.88\%, recall is 95.76\%, and F1-score is 96.32\%) (see Table \ref{tmperformance} in Appendix \ref{tmappdx} 
for full results).

In summary, the answer to \textbf{RQ1} is that our CRC and TM models effectively detect S\&P issues in user reviews with a micro-averaged F1 score of 98\% and 96\%, respectively.

\section{Results}
\subsection{Prevalence of S\&P concerns}

We now address our second research question, \textbf{RQ2}, regarding the prevalence (indicated by the proportion of reviews) of S\&P concerns in IoT product reviews. Task 1 of our CRC model classifies 4,896 out of 91,149 reviews (5.37\%) from the wild dataset to have S\&P concerns. To compare these concerns across device category $c$, we calculated the \textit{concern ratio} ($|C_c|/|R_c|$), where $|C_c|$ is the count of reviews with concerns and $|R_c|$ is the total reviews in category $c$. We check for significant differences in concern ratio with the Chi-squared test for proportions, followed by pairwise comparison with Bonferroni correction to control family-wide error rate.
The Chi-squared test for proportion (Table~\ref{tab:chisquaredtable} in Appendix \ref{pairwisechi}, middle) assesses whether the concern ratios are consistent across the device categories. The test yielded a $p<2.2e^{-16}$, indicating a highly significant difference in concern ratios across the categories. Pairwise comparisons (Table~\ref{tab:chisquaredtable}, right) showed significant differences between each pair of categories (all $p<0.001$), rejecting the null hypothesis and confirming that the concern ratios are not the same across trackers, speakers, and cameras. While the pairwise comparison shows significant inequality between the concern ratios of trackers and speakers, their numerical values are relatively similar (trackers = 2.11\% vs. speakers = 2.64\%). In contrast, cameras have a significantly higher concern ratio of 10.07\%, surpassing that of trackers and speakers. 
Prior work found 0.14\% S\&P reviews for fitness trackers~\cite{van2020pets}, 2.04\% for smart speakers~\cite{fruchter2018consumer}, and 5.9\% for cameras~\cite{vetrivelexamining}. Thus our classifier identifies 15× more S\&P reviews for fitness trackers, 29\% more for speakers, and 70\% more for cameras. This is likely due to higher recall compared to prior solutions (Table~\ref{tab:performance}).

We checked for correlations between the S\&P reviews and the star ratings by considering the counts of S\&P concerned and non-concerned reviews across different star ratings, and conducting a point-biserial correlation test. Results are shown in Table~\ref{pbcorrelationtest} in Appendix \ref{pbtest}. Comparing correlation coefficients ($r_{pb}$) category-wise and overall, we see no correlation in speakers, however we do observe low negative correlation otherwise.
This confirms a negative correlation between S\&P concerns and review ratings, indicating that such concerns likely contribute to lower user satisfaction, as noted in prior work~\cite{vetrivelexamining}.


\subsection{Types of S\&P Concerns}
\label{sec:concerns}

\begin{figure*}[!t]
\centering
\includegraphics[width=1.0\textwidth]{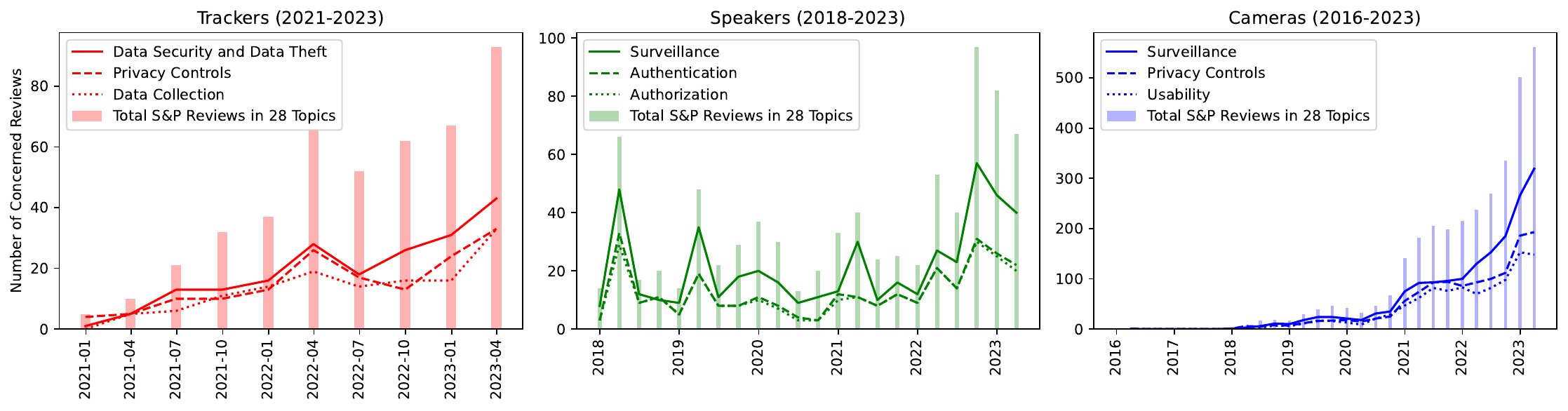}
\caption{Quarterly trends of the top three S\&P themes and unique review counts with S\&P concerns across trackers, speakers, and cameras. Each review is counted once, even if it covers multiple themes.}
\label{timeline}
\vspace{-16pt}
\end{figure*}

We now look at \textbf{RQ3}, which asks what the S\&P concerns are and how they rank. In Table~\ref{tab:theme-distro} we rank all the concerns for each device, showing the top concerns as well as lesser visible but important issues, and the relative level of concerns across these issues. Our longitudinal data allows us to view the evolution of reviews for top themes per device over multiple years, as shown in Figure~\ref{timeline}. The plot displays absolute values rather than normalized proportions (which is around 5\% per year) since our analysis focuses on trends in total user contributions rather than relative growth (see discussion below). We include the total unique S\&P review counts as separate bars to indicate likely increasing product adoption (the amount of historical data for each device type was influenced by our process for searching for the top devices in a category).

\begin{table*}[!ht]
\resizebox{1.0\textwidth}{!}{
\begin{tabular}{lllllllll}
\toprule
\multicolumn{3}{c|}{\textbf{Trackers}}                                                                      & \multicolumn{3}{c|}{\textbf{Speakers}}                                                                     & \multicolumn{3}{c}{\textbf{Cameras}}                                               \\ \midrule
\textbf{High-level Theme}                            & \textbf{\#} & \multicolumn{1}{r|}{\textbf{\%}} & \textbf{High-level Theme}                            & \textbf{\#} & \multicolumn{1}{r|}{\textbf{\%}} & \textbf{High-level Theme}                            & \textbf{\#} & \textbf{\%} \\ \midrule
1. data security and data theft                      & 215        & \multicolumn{1}{r|}{12.47}       & 1. surveillance                                      & 501        & \multicolumn{1}{r|}{17.32}       & 1. surveillance                                      & 1914       & 16.56       \\
2. privacy controls                                  & 175        & \multicolumn{1}{r|}{10.15}       & 2. authentication                                    & 302        & \multicolumn{1}{r|}{10.44}       & 2. privacy controls                                  & 1347       & 11.66       \\
3. data collection                                   & 155        & \multicolumn{1}{r|}{8.99}        & 3. authorization                                     & 288        & \multicolumn{1}{r|}{9.96}        & 3. usability                                         & 1116       & 9.66        \\
4. usability                                         & 116        & \multicolumn{1}{r|}{6.73}        & 4. access control                                    & 232        & \multicolumn{1}{r|}{8.02}        & 4. data security and data theft                      & 992        & 8.58        \\
5. authentication                                    & 109        & \multicolumn{1}{r|}{6.32}        & 5. privacy controls                                  & 218        & \multicolumn{1}{r|}{7.54}        & 5. access control                                    & 799        & 6.91        \\
6. policies and law                                  & 99         & \multicolumn{1}{r|}{5.74}        & 6. data security and data theft                      & 212        & \multicolumn{1}{r|}{7.33}        & 6. authentication                                    & 701        & 6.07        \\
7. access control                                    & 98         & \multicolumn{1}{r|}{5.68}        & 7. usability                                         & 192        & \multicolumn{1}{r|}{6.64}        & 7. authorization                                     & 655        & 5.67        \\
8. authorization                                     & 92         & \multicolumn{1}{r|}{5.34}        & 8. data collection                                   & 137        & \multicolumn{1}{r|}{4.74}        & 8. trust and transparency                            & 642        & 5.56        \\
9. location tracking                                 & 88         & \multicolumn{1}{r|}{5.1}         & 9. secure communication                              & 119        & \multicolumn{1}{r|}{4.11}        & 9. policies and law                                  & 424        & 3.67        \\
10. surveillance                                     & 76         & \multicolumn{1}{r|}{4.41}        & 10. trust and transparency                           & 117        & \multicolumn{1}{r|}{4.05}        & 10. data management and storage                      & 400        & 3.46        \\
11. trust and transparency                           & 64         & \multicolumn{1}{r|}{3.71}        & 11. privacy ethics                                   & 75         & \multicolumn{1}{r|}{2.59}        & 11. general comments related to S\&P & 374        & 3.24        \\
12. data accuracy                                    & 54         & \multicolumn{1}{r|}{3.13}        & 12. general comments related to S\&P & 73         & \multicolumn{1}{r|}{2.52}        & 12. secure communication                             & 321        & 2.78        \\
13. general comments related to S\&P & 45         & \multicolumn{1}{r|}{2.61}        & 13. policies and law                                 & 70         & \multicolumn{1}{r|}{2.42}        & 13. data collection                                  & 316        & 2.73        \\
14. consent                                          & 44         & \multicolumn{1}{r|}{2.55}        & 14. personalized advertising                         & 64         & \multicolumn{1}{r|}{2.21}        & 14. security vulnerabilities                         & 233        & 2.02        \\
15. data sharing                                     & 41         & \multicolumn{1}{r|}{2.38}        & 15. data sharing                                     & 49         & \multicolumn{1}{r|}{1.69}        & 15. software and firmware updates                    & 207        & 1.79        \\
16. personalized advertising                         & 41         & \multicolumn{1}{r|}{2.38}        & 16. security vulnerabilities                         & 47         & \multicolumn{1}{r|}{1.63}        & 16. personalized advertising                         & 188        & 1.63        \\
17. software and firmware updates                    & 39         & \multicolumn{1}{r|}{2.26}        & 17. software and firmware updates                    & 43         & \multicolumn{1}{r|}{1.49}        & 17. privacy ethics                                   & 146        & 1.26        \\
18. secure communication                             & 37         & \multicolumn{1}{r|}{2.15}        & 18. consent                                          & 33         & \multicolumn{1}{r|}{1.14}        & 18. data deletion                                    & 145        & 1.25        \\
19. data management and storage                      & 27         & \multicolumn{1}{r|}{1.57}        & 19. data harms                                       & 23         & \multicolumn{1}{r|}{0.8}         & 19. availability                                     & 132        & 1.14        \\
20. data exposure                                    & 22         & \multicolumn{1}{r|}{1.28}        & 20. data deletion                                    & 22         & \multicolumn{1}{r|}{0.76}        & 20. data sharing                                     & 116        & 1           \\
21. security vulnerabilities                         & 21         & \multicolumn{1}{r|}{1.22}        & 21. data management and storage                      & 21         & \multicolumn{1}{r|}{0.73}        & 21. data accuracy                                    & 99         & 0.86        \\
22. data deletion                                    & 15         & \multicolumn{1}{r|}{0.87}        & 22. confidentiality                                  & 16         & \multicolumn{1}{r|}{0.55}        & 22. consent                                          & 98         & 0.85        \\
23. data harms                                       & 15         & \multicolumn{1}{r|}{0.87}        & 23. location tracking                                & 12         & \multicolumn{1}{r|}{0.41}        & 23. data harms                                       & 51         & 0.44        \\
24. privacy ethics                                   & 15         & \multicolumn{1}{r|}{0.87}        & 24. data accuracy                                    & 8          & \multicolumn{1}{r|}{0.28}        & 24. location tracking                                & 47         & 0.41        \\
25. availability                                     & 11         & \multicolumn{1}{r|}{0.64}        & 25. availability                                     & 7          & \multicolumn{1}{r|}{0.24}        & 25. data exposure                                    & 34         & 0.29        \\
26. confidentiality                                  & 10         & \multicolumn{1}{r|}{0.58}        & 26. data exposure                                    & 6          & \multicolumn{1}{r|}{0.21}        & 26. confidentiality                                  & 31         & 0.27        \\
27. anonymity                                        & 0          & \multicolumn{1}{r|}{0}           & 27. anonymity                                        & 4          & \multicolumn{1}{r|}{0.14}        & 27. anonymity                                        & 22         & 0.19        \\
28. data hiding                                        & 0          & \multicolumn{1}{r|}{0}           & 28. data hiding                                        & 1          & \multicolumn{1}{r|}{0.03}        & 28. data hiding                                        & 7         & 0.06        \\ \bottomrule
\end{tabular}}
\caption{Distribution of high-level S\&P themes (28 in total) across different device categories. The column \# denotes the number of reviews tagged with a high-level theme. The column \% denotes the overall percentage of their prevalence.}
\label{tab:theme-distro}
\vspace{-12pt}
\end{table*}

\paragraphb{Fitness Trackers' Top Issues}
Table~\ref{tab:theme-distro} shows that the top three issues for fitness trackers are {\tt Data Security and Data Theft}, {\tt Privacy Controls}, and {\tt  Data Collection}. In {\tt Data Security and Data Theft} reviews, users express significant worry about apps that require credit card information for ‘free’ trials, as they fear it can lead to reports of unauthorized charges and account breaches. 
One user writes:
\begin{myquote}
``Shortly after using this and giving it all the permissions it asked for,
there were mysterious charges on my debit card...''
\end{myquote}
In the reviews about {\tt Privacy Controls}, users express frustration with the limited privacy controls to customize the apps or notifications (especially when they display sensitive data), and to restrict third-party data sharing. For example:
\begin{myquote}
``I hate that it shows what day of my menstrual cycle I’m on. How embarrassing ... no way to remove that''.    
\end{myquote}
We found a sizeable number of reviews about embarrassing notifications and an inability to control them. This indicates users are worried about their privacy in front of others in the same room, possibly due to shoulder surfing. In reviews labeled {\tt Data Collection}, users express concern that fitness trackers and their apps collect excessive personal data. These concerns have been reported in prior work, and we expected them to appear in our corpus. 

Our ranking of themes reveals that a number of issues are more frequent than location tracking.
This contrasts prior work in which location tracking for fitness trackers has been identified as a top 3 theme \cite{gabriele2020understanding}.This contrast may arise because prior studies are smaller in scale (e.g., 200 participants), while our automated analysis examines thousands of reviews and tens of topics.

Our temporal measurement in Figure~\ref{timeline} shows steady growth in the absolute number of reviews for the top 3 issues, suggesting that as device adoption increases, related concerns are rising too—indicating these issues remain inadequately addressed by the industry. 

\paragraphb{Smart Speakers' Top Issues}
The top issue for Smart Speakers is {\tt Surveillance}, comprising 17\% of the reviews. In these reviews about privacy, we see that users are deeply anxious about the ‘always-on’ nature of smart speakers, fearful of continuous conversation recording and data sharing with manufacturers, the government, or third parties. Concerns are heightened by unexpected activations without a clear wake word. An example review is:

\begin{myquote}
``You can believe the stories you hear about [PRODUCT] listening,
... talk about something and it would appear in your news feed
advertisements. Scary. Unplugged it.''
\end{myquote}
This review is interesting as it indicates that the user did more than complain --- they stopped using the device.

The privacy challenges of this sort have been identified before through surveys \cite{lutz2021privacy} and via experimentation \cite{dubois2020speakers}. One benefit of automated review analysis is that it enables longitudinal trend tracking. In Figure~\ref{timeline}, we see that the number of reviews about surveillance issues for smart speakers has remained fairly stable over a 5-year period. We examined the spikes in this plot and found that they mostly happen in January. We found many mentions of words like ``gift", ``holiday", and ``Christmas", suggesting these spikes are occurring after people have received holiday gifts.

The second and third issues are {\tt Authentication} and {\tt Authorization}. In these reviews, users tell stories about unauthorized access to user accounts and devices by people they view as scammers who impersonate others to gain remote access and steal personal information. For example:
\begin{myquote}
``I used the pop up virtual assistant during setup who remoted into my
computer, and installed a network security program. Two days later I got
8 calls about fraudulent charges on my account.''
\end{myquote}

The temporal trends for smart speakers (from Figure~\ref{timeline}) appear stable. Since the adoption is likely growing, this indicates a relative decrease in complaints about smart speakers. We can only speculate about this, either being due to growing learned helplessness or device improvement; this would be interesting to explore in future work.

\paragraphb{Camera's Top Issues}
As with smart speakers, {\tt Surveillance} is the top issue for cameras (17\% of S\&P reviews). Users fear unauthorized
surveillance by strangers, landlords, or even family members
without their knowledge or consent. The following review implies the user worries their landlord may have access to their video and knows when no one is home:
\begin{myquote}
``I surely hope to never find our landlord sneaking in when we are not
home as we heard he is nosy and goes in tenants apartments.''  
\end{myquote}
Park et al.~\cite{park2024well} have interviewed short term rental hosts about their IoT device usage, and identified that hosts/landlords often have competing needs and goals (e.g., protecting property) from protecting guests' privacy. Beyond landlords, we also saw reviews stating concerns about neighbors, friends, and family members accessing footage. 

In addition, many users worry that recordings are shared with governments and or third-party corporations.
\begin{myquote}
``I own a bunch of these cameras but will replace them and never buy a [COMPANY] products as long as they give MY FOOTAGE to authorities without consent or warrant. Do not enable this behavior by supporting [COMPANY].''
\end{myquote}

In the {\tt Privacy Controls} theme, users complain about not having the right level of control for data sharing as described in the review below: 
\begin{myquote}
``I purchased 4 minicams but I’m not happy. The cameras lack basic
features like not being able to share access ...
without supplying them your user credentials.''
\end{myquote}

In {\tt Usability}, the third top theme, users complain about the lagging performance of cameras, causing missed events and delayed notifications, as well as difficulties with the user interfaces in accessing live feeds and challenges managing multiple cameras. One user writes:
\begin{myquote}
``When you get alerts on your phone about motion detection, there’s no
way to disable individual cameras - you can disable all or enable all.''  
\end{myquote}
In Figure~\ref{timeline}, we see a stark growth of reviews discussing these three topics. The growth of surveillance is noteworthy, as this issue has been known for a while, and thus, we hypothesize that the problem is not really getting addressed. 

Many of the remaining dominant issues for speakers and cameras are related to multi-user households and shared account management difficulties. We thus discuss multi-user environments separately in a later section.

\subsection{Cross Device Insights}

\paragraphb{Customer Loss} Customer retention is a critical challenge for developers. Having found some reviews that discuss replacing devices or stopping to use them, we investigated more broadly if security and privacy issues were causing users to abandon apps or devices. We utilized GPT-5 (see the prompt in Appendix \ref{gpt5}) to label the reviews. This analysis identified 321 out of 4896 concern reviews, representing 6.6\% of the total across all device types. We then manually verified these reviews to confirm that users had indeed stopped using the device or service. Our straightforward prompt and manual review likely led to an underestimation of this behavior. 
For example, one user says:
\begin{myquote}
"Because it wants complete access to everything on my phone, such as phone numbers, messages, photos, contacts and "other files", I returned this device."
\end{myquote}
From these reviews, we extracted 15 recurring themes, indicating that a variety of issues contributed to customer loss. The most prominent reasons included {\tt Surveillance}, {\tt Data Security and Theft} and {\tt Privacy Controls}. However, we also found that reviews highlighting {\tt Trust and Transparency}, {\tt Policy and Law}, and {\tt Privacy Ethics} showcase when user dissatisfaction can lead to abandonment.


\paragraphb{Seeking More Privacy Control} 
We found users writing about wanting more fine-grained controls across all device types and many themes, including {\tt Privacy Controls, Authentication, Data Sharing, Access Control, Surveillance}. For example, consider the {\tt Privacy Controls} issue. With fitness trackers, users want more control over notifications that reveal sensitive information, while speaker users are concerned about the inability to control the ``always-listening" feature, and camera users focus on limited control over motion detection zones and who has access to the stored video.



\paragraphb{Ethics and the Law}
Because our taxonomy enables the separation of reviews into many categories, we are able to look at issues beyond the top few to gain a deeper understanding of users' perspectives. For example, in the theme of {\tt Privacy Ethics}, we see comments indicating a loss of trust. One speaker user writes
\begin{myquote}
``Your internet will be automatically shared through [BRAND-NAME] devices with people near ur home. This policy is not customer centric... Really degrades trust. [COMPANY] should have notified us.''
\end{myquote}
Similarly, one camera user states 
\begin{myquote} 
``Let's be clear, this is not a technology limitation, it's a purposeful limitation to force you into a [COMPANY] subscription. This kind of crap is why we can't have nice things.''
\end{myquote}

The theme {\tt Policy and Law} is a top 10 issue for trackers and cameras, and top 13 for speakers, and addresses a more specific aspect of data collection. For both speakers and cameras, we see reviews in which users discuss their data being given to government or law enforcement agencies. This illustrates an awareness of this data-sharing practice. Interesting, we see a divergent set of views about this. One upset reviewer says about their speaker product: 
\begin{myquote}
``A clause in the fine print allows law-enforcement agencies to simply fill out a form with Amazon to invade your privacy ... did not get any notification ... do I see a way to opt-out to protect my right to privacy ... buyer beware!''
\end{myquote}
Another review for a camera product, however, espouses an opposing view, showing they see validity for preventing physical harm.

\begin{myquote}
``If we reasonably believe that we can prevent someone from dying or from suffering serious physical harm, we may provide information to a government agency — for example, in the case of bomb threats, school shootings, kidnappings, suicide prevention, and missing persons cases... Why not?''
\end{myquote}
Sharing these reviews with developers can aid them in improving product communications to clarify data sharing with governments or law enforcement.

\section{Multi-User/Device Interactions Use Case}
\label{sec:multiuser}

The previous section shared that many users write about challenges occurring due to multi-user and multi-device interactions. These issues can surface under the \texttt{Authorization} (one user can access another's account), \texttt{Privacy controls} (one user changes another's settings), or \texttt{Usability} (cannot figure out how to manage multiple users) themes. In a collaborative space, proper multi-user interaction is of paramount importance to ensure that the privacy of each user is not violated. In this section, we highlight four different types of multi-user and multi-device interaction issues identified in our review analysis. We focus on reviews for speakers and cameras, as very few such reviews exist for fitness trackers.

One problematic scenario is when a single device is shared by many people, and the device supports only one account. Many home security camera systems lack granular access controls, preventing users from assigning different permission levels to multiple users (such as controlling which stored videos they can see). Device owners are often forced to share full administrative credentials with individuals who only need limited viewing. For example,
\begin{myquote}
``The biggest issue is that you can only have one login per setup ... if you want someone else to have access to the camera, you have to share your login details. That person then also has full admin rights ..., and there's no way to see what anyone might do...''

\end{myquote}

This situation also limits notification capability since the camera cannot tailor alerts for specific users.

A second problematic interaction occurs when different user accounts and their personal devices are unintentionally linked due to using the same app/service within the same household. In one incident, one user had full visibility of another user's request to the voice assistant.
\begin{myquote}
``BEWARE!!! Your alexa app may be linked to other people's apps even if they are using their own account ... bought a second echo as a gift for my mother. I could see everything she requested on her echo and she could see what I requested on mine. She was logged into alexa through her own account and the same was true for me.''
\end{myquote}

In another example, multiple speakers are present, and the notifications from one device get announced on another device, creating unwanted privacy disturbances. This creates potential for unauthorized access and data sharing.
\begin{myquote}
``Thought my unit would be some what independent of my roommates unit. Not so. Have registered units several times with little success. ... Could not drop in without problems of a contact being added to both units or their friend coming in on me.''
\end{myquote}

In a third scenario, one user has multiple devices in multiple locations and wishes to manage them via a single account. For some devices, this requires multiple log-ins and separate accounts, creating inconvenience and security risks. 
\begin{myquote}
``And when I wanted to install cameras on my property and my rental I was told I would have to start a whole new [COMPANY] account AND get another subscription. Then it requires me to log in and out all day to keep a consistent eye on all of the properties.''
\end{myquote}

A fourth scenario is when camera usage generates concerns about bystander privacy, particularly in shared spaces where they can capture multiple users without their consent.  We uncovered such  practices among some consumers who use cameras to secretly monitor specific family members or tenants without their knowledge. 
\begin{myquote}
``I purchased these to spy on my stepson and his lazy girlfriend, they gonna stop getting into my stuff now because I can see everything they do even when we're gone.''
\end{myquote}
This underscores the need for more transparent and ethically designed camera systems to protect the privacy of all individuals in shared environments.

The challenges due to multi-user and multi-device interactions have been partly explored before. Lee et al. have focused on designing privacy-preserving multi-user access control systems for IoT environments~\cite{lee2024fluid}, but did not first survey users to identify their concerns or needs.   Of the four scenarios we shared earlier, scenarios 2 and 3 are reported here for the first time, while scenarios 1 \cite{park2023nobody,lee2024fluid} and 4 \cite{ahmad2020tangible,park2023nobody} have been reported previously. The study in  ~\cite{park2023nobody} is based on interviewing 32 smart home power users. In contrast, our observations are derived from thousands of reviews, from 52 speaker and camera devices, and gathered over a multi-year period. Thus not only do we confirm prior findings, but we demonstrate them on a much larger scale. 

\section{Discussion}
\label{sec:discussion}

\subsection{Generalizability on Unseen Devices}
To demonstrate the generalizability of the CRC and TM models beyond the training data, we collected Reddit posts related to device types not included in training (i.e., virtual reality devices). We collected 11,233 posts from three VR-related subreddits: r/VirtualReality (4,124 posts), r/Oculus (3,944 posts), and r/VisionPro (3,165 posts). Applying the CRC model, we identified subsets of posts discussing S\&P issues (200 from r/VirtualReality, 252 from r/Oculus, and 114 from r/VisionPro). To evaluate accuracy, we randomly sampled 50 posts (17 from r/VirtualReality, 17 from r/Oculus, and 16 from r/VisionPro), which two coders manually labeled to create ground-truth labels. Inter-rater reliability was high, Cohen’s $\kappa$ was 0.96, which is almost perfect, according to \cite{landis1977measurement}. 
 Using these finalized labels, our model achieved an accuracy of 98\%, indicating strong performance in unseen contexts. To further assess the generalizability of the TM model, we evaluated it on 100 randomly selected VR-related posts flagged for S\&P issues. Again, two coders independently annotated the posts, achieving high initial agreement; Cohen’s $\kappa$ was 0.71, which is substantial, according to \cite{landis1977measurement}; any remaining conflicting labels were resolved and finalized through collective discussion. The finalized ground-truth labels were then compared against the model’s outputs. The TM model achieved an accuracy of 88\%, providing strong evidence that it generalizes effectively across contexts.

\subsection{Recommendation}

\paragraphb{Q\&A Agent for E-commerce Consumers} 
This work offers valuable support for consumer decision-making. Since S\&P reviews make up less than 10\% of total reviews for many IoT products, manually browsing or keyword searches often fail to surface relevant concerns. While early stage chatbots like Amazon's Rufus can highlight reviews on specific features, they currently lack the ability to accurately capture S\&P issues. Our approach fills this gap by automatically identifying S\&P-relevant reviews and generating both fine-grained and high-level thematic tags. These insights help surface meaningful comparisons across devices, enabling more informed purchases. A promising direction for future is to explore how to integrate S\&P summaries into GenAI assistants (e.g., Rufus) in a user-centric way that educates consumers while preserving review sentiment.

\paragraphb{Guidance for Product Vendors}
Stakeholders such as product vendors and developers can leverage our methodology to surface and summarize S\&P concerns from user reviews within development tools. Amazon sellers, for instance, already access customer feedback through the ``Feedback Manager'' portal. Our approach can enhance this by extracting and quantifying S\&P issues, helping prioritize concerns during feature planning.

Two key categories of developer takeaways emerged. First, users consistently seek more granular control—both over recipients (e.g., household members, landlords, third parties, law enforcement) and over specific content (e.g., selectively sharing audio/video footage or organizing it by day/device). Lack of such control often leads to user frustration and device abandonment.

Second, there is a clear need for better support in multi-user and multi-device scenarios. We identify four key challenges: (1) single devices that are shared but are lacking in multi-account support, (2) shared devices enabling unintended visibility on non-shared devices, (3) demand for unified control of devices across multiple homes, and (4) camera misuse for surveillance. Addressing these scenarios is crucial to improving user trust and retention.

Furthermore, we looked across all three device categories and who are the manufacturers of the devices being reviewed. We consistently found that the devices discussed in the reviews come from a small number of manufacturers (between 2-4) per device type.  As shown in \cite{peddintiimc2019}, surfacing nudges to developers showing that their competitors are not doing something that they do, and is unliked by users, can motivate change.

\paragraphb{Ethical Considerations of Using User Reviews}
Amazon currently generates AI-based summaries of user reviews without requiring explicit user opt-in or consent—submitting a review is considered implicit consent. Our envisioned use of assistive agents for both consumers and sellers aligns with Amazon's existing terms of service. However, we strongly advocate for greater transparency: a clear and explicit statement on the web portal regarding the use of user-generated data by AI is essential.

\paragraphb{Limitations/Biases}
Our work has a few limitations. First, we focus our analysis on reviews from only three device categories. However, given that Amazon is a top IoT device seller, we examined over 70 devices types, and that our models and taxonomy are not device-dependent, we believe our approach can generalize to other products too. Second, our analysis is limited to reviews in English. However, given that state-of-the-art LLMs are inherently multilingual, 
our approach can be readily extended to non-English reviews. Third, we focused on expressed user concerns without assessing the technical accuracy of users' perceptions.
Fourth, our analysis could be impacted by fake reviews on Amazon~\cite{he2022market}. Since we do not collect any user-specific metadata (such as usernames, even if publicly accessible), it is really challenging for us to identify fake reviews just by analyzing the review texts. We rely on Amazon’s mechanisms~\footnote{\url{https://www.amazon.com/gp/help/customer/display.html?nodeId=G8CXDFT9GLRRSV3G}} for removing these fake
reviews and banning abusive accounts, but acknowledge that some fake reviews may have trickled into our dataset. Lastly, sociodemographic characteristics of users and their biases may influence who submits reviews on Amazon, and these biases also impact our results.

\section{Conclusion}
We propose a modern approach to identifying security and privacy concerns related to IoT devices by prompt engineering and fine-tuning GPT-3.5-Turbo. Compared to prior work, this approach detects more issues and groups them into clear themes. Analyzing over 70 products across years, we find persistent issues, highlighting limited industry progress. The study particularly highlights two major challenges: users' demand for more precise control over their data, and the complexity of managing multi-user and multi-device setups, both of which must be carefully addressed in future design and development efforts.

\section{Acknowledgments}
We thank our anonymous reviewers for their valuable feedback. The opinions, findings, conclusions, or recommendations expressed in this work are those of the authors and do not necessarily reflect the views of the participating organizations.

{\small
\bibliography{aaai25}}

\pagebreak
\newpage
\clearpage
\appendix

\section*{Appendix}






\section{Algorithms to Generate Helpful Examples for The CRC and The TM Models} \label{sec:algos}

\begin{algorithm}[h]
\caption{Algorithm for generating two helpful examples to be shown at the prompt for the CRC model}\label{euclid}
\textbf{Input}: $t$ and $T$, where\\ \hspace*{\algorithmicindent} \hspace*{\algorithmicindent} $t = $ An unlabeled text passed to the CRC model\\ \hspace*{\algorithmicindent} \hspace*{\algorithmicindent}
    $T = $ The training dataset
    \\
\textbf{Output}: $t_1$, $t_2$, $Label(t_{1})$ and $Label(t_{2})$\\ 
\begin{algorithmic}[1] 

    \STATE $T \gets t \cup T$
    \STATE $L \gets Label(t) \cup Label(T)$ 
    \STATE $B \gets BertEmbed(T)$
    \STATE $C \gets CosineSimilarityMatrix(B)$
    \STATE $c \gets C[t]$
    \STATE $maxSim = - \infty$
    \STATE $maxSimIdx = -1 $
    \FOR {$i = 1$ to $|c|$}
    \IF {$c[i] \neq 1$ and $c[i] > maxSim$} 
    \STATE $maxSim \gets c[i] $
    \STATE $maxSimIdx \gets i$
    \ENDIF
    \ENDFOR
    \STATE $secondMaxSim = - \infty$
    \STATE $secondMaxSimIdx = -1 $
    \FOR {$i = 1$ to $|c|$}
    \IF {$c[i] \neq 1$ and $c[i] < maxSim$} 
        \IF {$c[i] > secondMaxSim$ and $Label(T[i]) \neq Label(T[maxSimIdx])$ } 
        \STATE $secondMaxSim \gets c[i] $
        \STATE $secondMaxSimIdx \gets i$
        \ENDIF
    \ENDIF
    \ENDFOR
    \STATE $t_{1} \gets T[maxSimIdx]$
    \STATE $Label(t_{1}) \gets L[maxSimIdx]$
    \STATE $t_{2} \gets T[secondMaxSimIdx]$
    \STATE $Label(t_{2}) \gets L[secondMaxSimIdx]$
    \STATE \textbf{return} $t_{1}, t_{2}, Label(t_{1}), Label(t_{2})$.
\end{algorithmic}
\end{algorithm}
\vfill\eject

\begin{algorithm}[!h]
\caption{Algorithm for generating five helpful examples to be shown at the prompt for the TM model}
\label{euclidtm}
\textbf{Input}: $t$ and $T$, where\\ \hspace*{\algorithmicindent} \hspace*{\algorithmicindent} $t = $ An unlabeled text passed to the TM model\\ \hspace*{\algorithmicindent} \hspace*{\algorithmicindent}
    $T = $ The training dataset
    \\
\textbf{Output}: $t_{1} - t_{5}$, $Label(t_{1})-Label(t_{5})$\\
\begin{algorithmic}[1] 
    \STATE $T \gets t \cup T$
    \STATE $L \gets Label(t) \cup Label(T)$
    \STATE $E \gets TFIDF(T)$
    \STATE $C \gets CosineSimilarityMatrix(E)$
    \STATE $c \gets C[t]$
    \STATE $Sort(c, decending)$
    \STATE $ArgSort(T) w.r.t$ $c$
    \STATE $ArgSort(L) w.r.t$ $c$
    \STATE $t_{1} \gets null $
    \STATE $t_{2} \gets null $
    \STATE $t_{3} \gets null $
    \STATE $t_{4} \gets null $
    \STATE $t_{5} \gets null $
    \FOR {$i = 2$ to $6$} 
    \STATE $t_{i-1} \gets T[i-1] $
    \ENDFOR
    \STATE $Label(t_{1}) \gets Label[1] $
    \STATE $Label(t_{2}) \gets Label[2] $
    \STATE $Label(t_{3}) \gets Label[3] $
    \STATE $Label(t_{4}) \gets Label[4] $
    \STATE $Label(t_{5}) \gets Label[5] $
    \STATE \textbf{return} $t_{1}, t_{2}, t_{3}, t_{4}, t_{5}$,\\ $Label(t_{1}), Label(t_{2}), Label(t_{3}), Label(t_{4}), Label(t_{5})$.
\end{algorithmic}
\end{algorithm}

\newpage
\onecolumn
\section{The CRC Model}

\subsection{The CRC prompt with an actual example from the training dataset}
\label{crctemplateexample}
\begin{itemize}
    \item  You are a Large Language Model for classifying, rationalizing, and categorizing security and privacy concerns in text.
    \item  Given the text T: \textit{\textbf{I bought this with no intention of utilizing the subscription and doing only local storage. When I first bought the product and set it all up, everything worked great and it would allow me to get alerts when there was motion and would shoot a little 10 second video so I could review it later. About 3 months or so later, that got moved behind the subscription paywall as well. Now I can only use them as live feed as any other service is blocked without paying them extra. Only buy Blink if you’re planning to buy the subscription as well
}}

    \item I want you to perform three tasks on T:
    \begin{itemize}
        \item Task 1: Determine if text T explicitly mentions any security or privacy concerns. Respond only with a Yes or a No.
        \item Task 2: Provide a brief explanation as to why your response is as such for Task 1.
        \item Task 3: If the response for Task 1 was a Yes, categorize the explicit concerns mentioned in text T with themes related to security and privacy.
    \end{itemize}
    
    \item Here are two examples that might be helpful:
    \begin{itemize}
        \item Example text A: \textit{\textbf{cannot use their app now, without giving them your phone number. useless cameras sitting on my pile of garage sale items. they work okay, if you can get the app to work. app and cameras worked fine for a few months, now i cannot use them as the app has an apparently new requirement that i give them a phone number so they can send me a text - on a landline. email was fine for this, now the app won’t allow getting past the page that requires this. customer service says to tap on an icon that does not appear on this page - just “give us your phone number”. done with blink, will find another camera company that is more oriented toward the customer}}
        \item Ideal response for A:\\\textit{\textbf{Task 1: Yes\\Task 2: The text raises security and privacy concerns because the app is now requiring the user to provide their phone number, which may be unnecessary for the functionality of the app. This raises questions about why the company needs access to the user's phone number and how they will use it. Additionally, the mention of the app not allowing the user to proceed without providing the phone number suggests a lack of transparency and control over personal information.\\Task 3: requirement of a phone number to get access to the app, lack of transparency in why a phone number is required to use the app}}
        \item Example text B: \textit{\textbf{my biggest complaint with these cameras having to tap the screenshot to update the picture. i also hate that you have to wait before loading into a view of another camera. it will prompt “live view in progress”never had that on my zmodo.
}}
        \item Ideal response for B:\\
        \textit{\textbf{Task 1: No\\Task 2: The text provided does not explicitly mention any security or privacy concerns. The complaints expressed are primarily related to the functionality and user experience of the cameras, such as the need to tap the screenshot to update the picture and the delay in loading into a view of another camera. These complaints do not directly pertain to security or privacy.\\Task 3: N/A
}}
    \end{itemize}
\item The following goes in the assistant prompt during training:\\
\textit{\textbf{Task 1: Yes\\
Task 2: The text raises concerns about the subscription-based model of the product. The user mentions that certain features, such as motion alerts and video recording, were initially available without a subscription but later moved behind a paywall. This raises questions about the security and privacy of the user's data. It suggests that the company may be limiting access to essential security features unless the user pays for a subscription, which could potentially compromise the user's security and privacy.\\
Task 3: Concerns about the subscription model, Limited access to security features, Potential data collection and privacy implications}}
\end{itemize}

\begin{table*}[!h]
\centering
\resizebox{1.0\textwidth}{!}{
\begin{tabular}{@{}lrrllrrrrrr@{}}
\toprule
\multicolumn{3}{c}{\textbf{Metric to Evaluate Classification}}         & \multicolumn{8}{c}{\textbf{Metric to Evaluate Rationale and Low-level Issue Generation}}                                                                                                                     \\ \midrule
\multirow{2}{*}{\textbf{Metric}} & \multicolumn{2}{c}{\textbf{Task 1}} & \multirow{2}{*}{\textbf{Metric}} & \multirow{2}{*}{\textbf{Objective}}     & \multicolumn{2}{c}{\textbf{Task 2}} & \multicolumn{2}{c}{\textbf{Task 3 = "N/A"}} & \multicolumn{2}{c}{\textbf{Task 3 = Other}} \\ \cmidrule(lr){2-3} \cmidrule(l){6-11} 
                                 & \textbf{Valid}    & \textbf{Test}   &                                  &                                         & \textbf{Valid}    & \textbf{Test}   & \textbf{Valid}        & \textbf{Test}       & \textbf{Valid}        & \textbf{Test}       \\ \midrule
Precision                        & 97.8\%            & 88.7\%          & ROUGE-L                          & \multirow{2}{*}{Overlapping Similarity} & 0.531             & 0.739           & 0.9793                & 0.9947              & 0.508                 & 0.6732              \\
Recall                           & 97.8\%            & 93.4\%          & METEOR                           &                                         & 0.5953            & 0.8119          & 0.9805                & 0.9893              & 0.6169                & 0.7462              \\ \cmidrule(l){4-11} 
F1-score                         & 97.8\%            & 91.0\%          & BERTScore                        & Semantic Similarity                     & 0.9395            & 0.9731          & 0.996                 & 0.999               & 0.936                 & 0.9416              \\ \bottomrule
\end{tabular}}
\caption{Performance of the CRC model for the three tasks. Task 3 = ``N/A", when Task 1 = ``No".}
\label{performance}
\vspace{24pt}
\end{table*}

%
%
%
%

%
%
%
%


\newpage

\section{The TM Model}\label{tmappdx}

\subsection{The Prompt for the TM Model}
\begin{figure}[!h]
\fbox{
\footnotesize
\begin{minipage}{.95\columnwidth}
------ System Prompt ------\\
You are a Large Language Model for mapping a low-level issue to one or more high-level themes.\\
------ User Prompt ------\\
I have $x$ , a low-level issue related to security and privacy:
\underline{\texttt{$x$, an individual Task 3 response}}\\
I also have $Y$, a list of high-level themes related to security and privacy:
\texttt{$Y$, taxonomies with their definitions}
\\
Your task is to map $x$ to one or more $y$'s from $Y$.\\
Respond in the form: $x \rightarrow y_{1}$ if $x$ should be mapped to just one $y$.\\
Respond in the form: $x \rightarrow y_{1}, y_{2}, ..., y_{n}$ if $x$ should be mapped to more than one $y$.\\
Here are some examples that might be helpful.\\
E.g., low-level theme A: \underline{\texttt{$t_{1}$}}; Response for A: \underline{\texttt{$Label(t_{1})$}}\\
E.g., low-level theme B: \underline{\texttt{$t_{2}$}}; Response for B: \underline{\texttt{$Label(t_{2})$}}\\
E.g., low-level theme C: \underline{\texttt{$t_{3}$}}; Response for C: \underline{\texttt{$Label(t_{3})$}}\\
E.g., low-level theme D: \underline{\texttt{$t_{4}$}}; Response for D: \underline{\texttt{$Label(t_{4})$}}\\
E.g., low-level theme E: \underline{\texttt{$t_{5}$}}; Response for E: \underline{\texttt{$Label(t_{5})$}}
\end{minipage}}
\caption{The prompt for the TM model}
\label{fg:tmpromptengineering}
\end{figure}

\subsection{The TM prompt with an actual example}\label{tmactualexample}
\begin{itemize}
    \item You are a Large Language Model for mapping a low-level theme to one or more high-level themes.
    \item I have x , a low-level theme related to security and privacy: \textbf{\textit{password sharing as a violation of basic it security principles}}
    \item I also have Y, a list of high-level themes related to security and privacy:
    \\ \fbox{\parbox{0.4\textwidth}{\textbf{\texttt{$Y$, the list of high-level themes\\ with their definitions}
        }}}
    \item Your task is to map $x$ to one or more $y$-s.
    \item This means map $x$ to one or more appropriate $y$-s from $Y$.
    \begin{itemize}
        \item Respond in the form: $x->y_{1}$ if $x$ should be mapped to just one $y$.
        \item Respond in the form: $x->y_{1}, y_{2}, ..., y_{n}$ if $x$ should be mapped to more than one $y$.
    \end{itemize}
    \item Here are some examples that might be helpful.
    \begin{itemize}
      \item Example low-level theme A: \textbf{\textit{password security}}

      \item Ideal response for A: \textbf{\textit{password security $->$ authentication}}
      
      \item Example low-level theme B: \textbf{\textit{concerns about password security}} 
      
      \item Ideal response for B: \textbf{\textit{concerns about password security $->$ authentication}}
      
      \item Example low-level theme C: \textbf{\textit{privacy violation}}
      
      \item Ideal response for C: \textbf{\textit{privacy violation $->$ surveillance}}

      \item Example low-level theme D: \textbf{\textit{concerns about privacy violation}}
      \item Ideal response for D: \textbf{\textit{concerns about privacy violation $->$ surveillance}}
      
      \item Example low-level theme E: \textbf{\textit{potential violation of privacy zones}}
      
      \item Ideal response for E: \textbf{\textit{potential violation of privacy zones $->$ privacy controls}}
    \end{itemize}
      \item The following goes in the assistant prompt during training:\\
      \textbf{\textit{password sharing as a violation of basic it security principles $->$ authentication, data sharing}}
\end{itemize}

\subsection{The labeled data used to create the training dataset for this work}
\begin{table*}[!h]
\centering
\footnotesize
\begin{tabularx}{\textwidth}{@{} XXX @{}}
\toprule
 \textbf{Text} & \textbf{Label from \cite{protick2024unveiling}} & \textbf{Converted label based on our need}                                           \\ \midrule
GOT HACKED$<$-tc-$>$ I had one in my sons room because he has no sense of danger and it got hacked and they kept moving the camera and talking through the camera to my son when he was asleep                             & Label from the review labeling step

Coder 1= 1

Coder 2= 1

Label from the thematic coding step

Coder 1= Vignette of being hacked or someone taking control over the camera

Coder 2= Hacking concerns
                                                               & Task 1: Yes

Task 2: The text mentions that a camera in the son's room was hacked, indicating a security breach. The hackers were able to control the camera, move it, and even communicate through it, which raises concerns about privacy invasion and potential harm to the son's well-being.

Task 3: hacking incident, breach of security
 \\ \bottomrule

\end{tabularx}
\caption{Conversion of the label from existing work~\cite{protick2024unveiling} for our model training}
\label{need-crc}
\vspace{-12pt}
\end{table*}

\clearpage
\subsection{Performance of the TM model on the validation dataset}\label{tmvalidperformance}

\begin{table*}[!h]

\resizebox{1.\textwidth}{!}{
\begin{tabular}{llllll}
\toprule
\textbf{Taxonomy Term}                           & \textbf{Rows Annotated} & \textbf{Rows Predicted} & \textbf{Precision} & \textbf{Recall} & \textbf{F1-score} \\ \midrule
access control                                   & 23                      & 23                      & 1                  & 1               & 1                 \\
anonymity                                        & 0                       & 0                       & 0                  & 0               & 0                 \\
authentication                                   & 41                      & 41                      & 1                  & 1               & 1                 \\
authorization                                    & 23                      & 23                      & 1                  & 1               & 1                 \\
availability                                     & 3                       & 2                       & 1                  & 0.67            & 0.8               \\
confidentiality                                  & 3                       & 3                       & 1                  & 1               & 1                 \\
consent                                          & 13                      & 13                      & 1                  & 1               & 1                 \\
data accuracy                                    & 4                       & 4                       & 1                  & 1               & 1                 \\
data collection                                  & 32                      & 33                      & 0.94               & 0.97            & 0.95              \\
data deletion                                    & 4                       & 4                       & 1                  & 1               & 1                 \\
data exposure                                    & 6                       & 5                       & 1                  & 0.83            & 0.91              \\
data harms                                       & 3                       & 3                       & 1                  & 1               & 1                 \\
data hiding                                      & 2                       & 0                       & 0                  & 0               & 0                 \\
data management and storage                      & 14                      & 14                      & 1                  & 1               & 1                 \\
data security and data theft                     & 35                      & 41                      & 0.85               & 1               & 0.92              \\
data sharing                                     & 17                      & 18                      & 0.94               & 1               & 0.97              \\
general comments related to security and privacy & 21                      & 21                      & 0.95               & 0.95            & 0.95              \\
location tracking                                & 7                       & 7                       & 1                  & 1               & 1                 \\
personalized advertising                         & 6                       & 5                       & 1                  & 0.83            & 0.91              \\
policies and law                                 & 23                      & 23                      & 0.96               & 0.96            & 0.96              \\
privacy controls                                 & 56                      & 58                      & 0.95               & 0.98            & 0.96              \\
privacy ethics                                   & 35                      & 31                      & 1                  & 0.89            & 0.94              \\
secure communication                             & 5                       & 5                       & 1                  & 1               & 1                 \\
security vulnerabilities                         & 14                      & 12                      & 1                  & 0.86            & 0.92              \\
software and firmware updates                    & 1                       & 1                       & 1                  & 1               & 1                 \\
surveillance                                     & 92                      & 89                      & 1                  & 0.97            & 0.98              \\
trust and transparency                           & 19                      & 18                      & 0.94               & 0.89            & 0.92              \\
usability                                        & 17                      & 16                      & 0.94               & 0.88            & 0.91              \\ \midrule
Macro-averaged                                   & -                        &    -                     & 0.91               & 0.88            & 0.89              \\
Micro-averaged                                   & -                        &  -                       & 0.97               & 0.96            & 0.96              \\ \bottomrule
\end{tabular}}
\caption{Performance of the TM model for each taxonomy terms across 360 validation examples. `Rows Annotated' are rows of examples mapped to that term as ground truth `Rows Predicted' are rows of examples for which the TM model mapped that example with that taxonomy term. The macro-averaged calculates the precision, recall, and F1-score for each taxonomy term independently and then averages them. The micro-averaged aggregates the contribution of all classes before calculating the metric.}
\label{tmperformance}
\end{table*}




\newpage

\clearpage 
\newpage

\section{Pairwise Chi-square Test}\label{pairwisechi}
\begin{table*}[!ht]
\centering
\resizebox{.9\textwidth}{!}{
\begin{tabular}{@{}llll|llll|lll@{}}
\toprule
\multirow{2}{*}{\textbf{Category}} & \multicolumn{3}{l|}{\textbf{Concern ratio}} & \multicolumn{4}{l|}{\textbf{Test for equality of concern ratio}}              & \multicolumn{3}{l}{\textbf{Pairwise test}}      \\ \cmidrule(l){2-11} 
                                   & \textbf{$|C_{c}|$}  & \textbf{$|R_{c}|$} & \textbf{$|C_{c}/R_{c}|$} & \textbf{Test} & \textbf{Test Statistic} & \textbf{$df$} & \textbf{$p$-value}      & \textbf{$c_{i}$} & \textbf{$c_{j}$} & \textbf{$p$-value}    \\ \midrule
Trackers                                 & 505          & 23894       & 0.0211         & \multirow{3}{*}{Chi-squared}   & \multirow{3}{*}{$\chi^{2}$ = 2498.9}           & \multirow{3}{*}{2}           & \multirow{3}{*}{\textless{}2.2e-16***} & Trackers          & Speakers         & 2e-04***            \\
Speakers                                & 847          & 32069       & 0.0264         &               &                         &             &                       & Trackers          & Cameras        & \textless{}2e-16*** \\
Cameras                               & 3544         & 35186       & 0.1007         &               &                         &             &                       & Speakers         & Cameras        & \textless{}2e-16*** \\ \bottomrule
\end{tabular}}
\caption{At the left: Calculation of the concern ratio. In the middle: test for equality of proportions. $H_{0}:$ $|C_{c}|/|R_{c}|$ is the same in each $c$. $H_{A}:$ At least one $|C_{c}|/|R_{c}|$ for a $c$ is different from the others. At the right: Posthoc pairwise comparison of the proportions of all pairs of $|C_{c}|/|R_{c}|$'s.}
\label{tab:chisquaredtable}
\end{table*}

\section{Point Biserial Test of Correlation between S\&P concerns in a review and rating given by a user} \label{pbtest}
\begin{table*}[!ht]
\centering
\resizebox{0.9\textwidth}{!}{\begin{tabular}{l|l|l|lllll|lllll}
\toprule
\multirow{2}{*}{\textbf{Category}} & \multirow{2}{*}{\textbf{Concern}} & \multirow{2}{*}{\textbf{Total}} & \multicolumn{5}{l|}{\textbf{Rating}} & \multicolumn{5}{l}{\textbf{Point-biserial test}}                                                                                                                          \\ \cline{4-13} 
                                   &                                   &                                 & \multicolumn{1}{l|}{\textbf{1}} & \multicolumn{1}{l|}{\textbf{2}} & \multicolumn{1}{l|}{\textbf{3}} & \multicolumn{1}{l|}{\textbf{4}} & \textbf{5} & \multicolumn{1}{l|}{\textbf{$t$}} & \multicolumn{1}{l|}{\textbf{$df$}} & \multicolumn{1}{l|}{\textbf{$p$-value}}   & \multicolumn{1}{l|}{\textbf{$r_{pb}$}} & \textbf{Interpretation} \\ 
                                   \midrule
\multirow{2}{*}{Trackers}                                 & No                                & 23389                           & \multicolumn{1}{l|}{4684}       & \multicolumn{1}{l|}{2263}       & \multicolumn{1}{l|}{2676}       & \multicolumn{1}{l|}{3364}       & 10402      & \multicolumn{1}{l|}{-16.603}    & \multicolumn{1}{l|}{23892}       & \multicolumn{1}{l|}{\textless{}2.2e-16***} & \multicolumn{1}{l|}{-.11}         & Low corr. \cite{kuckartz2013statistik}               \\
                                   & Yes                               & 505                             & \multicolumn{1}{l|}{235}        & \multicolumn{1}{l|}{75}         & \multicolumn{1}{l|}{55}         & \multicolumn{1}{l|}{64}         & 76         & \multicolumn{1}{l|}{}           & \multicolumn{1}{l|}{}            & \multicolumn{1}{l|}{}                   & \multicolumn{1}{l|}{}             &                         \\ \hline
\multirow{2}{*}{Speakers}                                & No                                & 31222                           & \multicolumn{1}{l|}{1716}       & \multicolumn{1}{l|}{1176}       & \multicolumn{1}{l|}{2437}       & \multicolumn{1}{l|}{6421}       & 19472      & \multicolumn{1}{l|}{-10.801}    & \multicolumn{1}{l|}{32067}       & \multicolumn{1}{l|}{\textless{}2.2e-16***} & \multicolumn{1}{l|}{-.06}         & No corr. \cite{kuckartz2013statistik}                \\
                                   & Yes                               & 847                             & \multicolumn{1}{l|}{92}         & \multicolumn{1}{l|}{51}         & \multicolumn{1}{l|}{100}        & \multicolumn{1}{l|}{226}        & 378        & \multicolumn{1}{l|}{}           & \multicolumn{1}{l|}{}            & \multicolumn{1}{l|}{}                   & \multicolumn{1}{l|}{}             &                         \\ \hline
\multirow{2}{*}{Cameras}                               & No                                & 31642                           & \multicolumn{1}{l|}{4057}       & \multicolumn{1}{l|}{1912}       & \multicolumn{1}{l|}{2667}       & \multicolumn{1}{l|}{4408}       & 18598      & \multicolumn{1}{l|}{-35.504}    & \multicolumn{1}{l|}{35184}       & \multicolumn{1}{l|}{\textless{}2.2e-16***} & \multicolumn{1}{l|}{-.19}         & Low corr. \cite{kuckartz2013statistik}               \\
                                   & Yes                               & 3544                            & \multicolumn{1}{l|}{970}        & \multicolumn{1}{l|}{469}        & \multicolumn{1}{l|}{537}        & \multicolumn{1}{l|}{438}        & 1130       & \multicolumn{1}{l|}{}           & \multicolumn{1}{l|}{}            & \multicolumn{1}{l|}{}                   & \multicolumn{1}{l|}{}             &                         \\ \hline
\multirow{2}{*}{Overall}                                & No                                & 86253                           & \multicolumn{1}{l|}{10457}      & \multicolumn{1}{l|}{5351}       & \multicolumn{1}{l|}{7780}       & \multicolumn{1}{l|}{14193}      & 48972      & \multicolumn{1}{l|}{-40.168}    & \multicolumn{1}{l|}{91147}       & \multicolumn{1}{l|}{\textless{}2.2e-16***} & \multicolumn{1}{l|}{-.13}         & Low corr. \cite{kuckartz2013statistik}               \\ 
                                   & Yes                               & 4896                            & \multicolumn{1}{l|}{1297}       & \multicolumn{1}{l|}{595}        & \multicolumn{1}{l|}{692}        & \multicolumn{1}{l|}{728}        & 1584       & \multicolumn{1}{l|}{}           & \multicolumn{1}{l|}{}            & \multicolumn{1}{l|}{}                   & \multicolumn{1}{l|}{}             &                         \\ \bottomrule
\end{tabular}}
\caption{Point-biserial test between S\&P concerns in a review and the rating given by a user. $H_{0}:$ $\rho_{pb} = 0$ and $H_{A}:$ $\rho_{pb} \neq 0$. The significance code *** means $p < .001$.}
\label{pbcorrelationtest}
\vspace{-4pt}
\end{table*}

\section{Prompt used in GPT-5 to detect cases of app/device ‘Stop Using,’ ‘Replacing,’ or ‘Uninstalling.’}\label{gpt5}
\begin{itemize}
    \item You are a text classification assistant. Your task is to carefully read a piece of text (an Amazon product review) and do a classification task.
    \item Given the text T: [USER REVIEW]
    \item Classify whether the reviewer explicitly indicates one of the following actions:
    \begin{enumerate}
        \item The user uninstalled the product.
        \item The user replaced the product.
        \item The user stopped using the product.
        \item None of the above three actions.
    \end{enumerate}
    \item Always output exactly one of: `1. Uninstalled', `2. Replaced', `3. Stopped Using', or `4. None of them'. Do not add explanations or extra text.
    \item If a user uninstalled, replaced, or stopped using the product due solely to configuration issues rather than both confirmation and privacy concerns, label it 4. None of them.
    
\end{itemize}
\newpage






\end{document}